\documentclass[runningheads]{llncs}
\usepackage{graphicx}
\usepackage{amsmath}
\usepackage{graphicx}
\usepackage{xcolor}   
\usepackage{hyperref} 
\usepackage{csquotes}
\usepackage{subcaption}
\usepackage{multirow}
\usepackage[normalem]{ulem}
\usepackage[absolute,overlay]{textpos} 

\begin{document}
\title{LSSD: a Controlled Large JPEG Image Database for Deep-Learning-based Steganalysis \enquote{into the Wild}
}
\titlerunning{LSSD: a Controlled Large JPEG Image Database for DL-based Steganalysis}
%
\author{
Hugo Ruiz\inst{1}\orcidID{0000-0002-2806-2428} \and
Mehdi Yedroudj\inst{1}\orcidID{0000-0001-6404-3876} \and \\
Marc Chaumont\inst{1,2}\orcidID{0000-0002-4095-4410} \and
Frédéric Comby\inst{1}\orcidID{0000-0001-7157-4296} \and
Gérard Subsol\inst{1}\orcidID{0000-0002-7461-4932}
}
\authorrunning{H. Ruiz et al.}
%

\institute{
Research-Team ICAR, LIRMM, Univ. Montpellier, CNRS, France\\
\email{\{hugo.ruiz, mehdi.yedroudj, marc.chaumont, \\ frederic.comby, gerard.subsol\}@lirmm.fr}
\and University of N\^{i}mes, France}
%
\maketitle              
\begin{abstract}
For many years, the image databases used in steganalysis have been relatively small, i.e. about ten thousand images. This limits the diversity of images and thus prevents large-scale analysis of steganalysis algorithms. 

In this paper, we describe a large JPEG database composed of 2 million colour and grey-scale images. This database, named LSSD for Large Scale Steganalysis Database, was obtained thanks to the intensive use of \enquote{controlled} development procedures. LSSD has been made publicly available, and we aspire it could be used by the steganalysis community for large-scale experiments.

We introduce the pipeline used for building various image database versions. We detail the general methodology that can be used to redevelop the entire database and increase even more the diversity. We also discuss computational cost and storage cost in order to develop images. 

\end{abstract}
%
%
%
	\section{Introduction}
	\label{sec:introduction}
\begin{textblock*}{122mm}(4.7cm,1cm)
{\footnotesize \textnormal{
\textit{ICPR'2021, International Conference on Pattern Recognition, MMForWILD'2021, Workshop on MultiMedia FORensics in the WILD, Lecture Notes in Computer Science, LNCS, Springer. January 10-15, 2021, Virtual Conference due to Covid (formerly Milan, Italy). Version of December 2020.\vspace{+0.7cm}\\}}}
\end{textblock*}

Steganography is the art of hiding information in an non suspicious medium so that the very existence of the hidden information is statistically undetectable from unaware individuals. Conversely, steganalysis is the art of detecting the presence of hidden data in such supports \cite{Fridrich2009_Book}. JPEG images are attractive supports since they are massively used in cameras and mobile phones and in all media of communication on the Internet and social networks. In this paper, we will then focus on  steganography and steganalysis in JPEG images.

In 2015, steganalysis using Deep Learning techniques emerged \cite{Chaumont2020}, and nowadays, they are considered as the most efficient way to detect stego images (i.e. images which contain a hidden message). 
Moreover, GPU computation capabilities increase regularly, which ensures faster computing speed which reduces learning time. So, performances of steganalysis methods based on Deep Learning have significantly improved.

Nevertheless, in most cases, the performance of steganalysis based on Deep Learning depends on the size of the learning image database. To a certain extent, the larger the database is, the better the results are \cite{Yedroudj2018_DatabaseAugmentation} \cite{Ruiz2021_Scaling}. Thus, increasing the size of the learning database generally improves performance while increasing the diversity of the examples. 
 
Currently, the most significant database used in steganalysis by Deep Learning is made of one million JPEG images \cite{Zeng2017_Millions} excerpted from the ImageNet database, which contains more than 14 million images. 

That said, databases created in "controlled" conditions, that is to say, with the full knowledge of the creation process of the images so that the development is repeatable, are not very big in comparison. Indeed, we can mention, as a \enquote{controlled} database the BOSS database \cite{Bas2011-BOSS} with a size of only 10,000 images and the Alaska \#2 database, with a size of 80,000 images \cite{Cogranne2020_Alaska2}.

In this paper, we present the "controlled" {\it Large Scale Steganalysis Database} (LSSD) which is a public JPEG image database, made of 2 million images, with a colour version and a grey-scale version, and which was created for the research community working on steganalysis.

One important aspect when creating an image database for steganalysis is to have diversity to get closer to reality \cite{Ker2013_RealWorld}. This \enquote{diversity} mainly depends on the ISO and the \enquote{development} process of the RAW image that is captured by the camera sensor. ISO is a measure of the sensitivity to light of the image sensor and if available, can be notified in the metadata associated with the JPEG image, i.e. in the EXIF metadata.

As in analogical photography, the \enquote{development} process consists in applying image processing operations (demosaicing, gamma correction, blur, colour balance, compression) in order to transform the RAW image into a viewable image in a standard format. The RAW image, when the camera is made of a colour filter array (CFA) of type Bayer filter (which is majority the case), is a unique 2D matrix containing 50\% green, 25\% red and 25\% blue. In order to \enquote{control} this diversity, it is possible to tune the different development parameters, without modifying the ISO parameters, and get different developments and then different JPEG versions of the same RAW image. Thus, if we want to increase the size of the database to more than 10 million images, it is necessary to implement a well-thought-out procedure to automate the \enquote{controlled} generation, to optimize the processing time as well as the storage volume.

By controlling the development, it becomes possible to get large databases which can be used for the learning or the test phases of Deep-Learning based steganalysis algorithms. We can then conduct an objective and repeatable evaluation of the performances of these algorithms. In particular, it will allow researchers to work on one of the major challenges in the steganalysis field: the \enquote{Cover-Source Mismatch}. Cover-Source Mismatch (CSM) is a phenomenon that occurs when the training set and the test set come from two different sources, causing bias in the Deep-Learning learning phase and resulting in bad results in the test phase.

In Section \ref{sec:creation}, we detail the whole development procedure that is used for the generation of the LSSD database. In Section \ref{sec:steganalysis}, we explain how to use the LSSD Database to create a learning set and a test set for Deep-Learning steganalysis applications. We also emphasize the problem of computational and storage cost for the creation of those sets.

\section{A \enquote{controlled} procedure to get a JPEG image}
\label{sec:creation}
	\subsection{RAW image sources}
	\label{ssec:raw}
To build a consistent \enquote{controlled} base, we chose to gather a maximum of RAW images, i.e. which are composed of the original sensor data. More precisely, we use the file that contains the RAW data of the sensor before any lossy compression (JPEG for example), and before any transformation required for its visualization on screen. It is important to note that each manufacturer adapts the data format to its hardware and then that we can find many formats (e.g., .dng, .cr2, .nef\ldots). 

At the contrary of JPEG images, RAW images are extremely rare on the Internet because they are large files and used by very few people. The size of a RAW image is usually around $3,000 \times 5,000$ pixels. Since data are in \enquote{raw} format, it represents a lot of information to store. It is therefore rare that Web sites, even those specialized in photography, dedicate a specific storage space to this kind of images. 

\begin{table}[htbp]
	\centering
	\caption{Number of images and devices used in each database.}
	\label{tab:intro:database_origins}
	\begin{tabular}{|l|c|c|}
		\hline
		\textbf{Database} & \textbf{number of } & \textbf{number of}\\
		\textbf{Name} & \textbf{images} & \textbf{devices}\\
		\hline
		\hline
		ALASKA2 \textsuperscript{\ref{fn:base:alaska2}} & 80,005 & 40\\
		\hline
		BOSS \textsuperscript{\ref{fn:base:boss}} & 10,000 & 7\\
		\hline
		Stego App DB \textsuperscript{\ref{fn:base:stegoapp}} & 24.120 & 26\\
		\hline
		Wesaturate \textsuperscript{\ref{fn:base:wesaturate}} & 3.648 & /\\
		\hline
		RAISE \textsuperscript{\ref{fn:base:raise}} & 8,156 & 3\\
		\hline
		Dresden \textsuperscript{\ref{fn:base:dresden}} & 1,491 & 73 (25 different models)\\
		\hline
		\hline
		\textbf{Total} & \textbf{127,420} & \textbf{101}\\
		\hline
	\end{tabular}
\end{table}

\begin{figure}[htbp]
	\centering

	\includegraphics[scale=0.50]{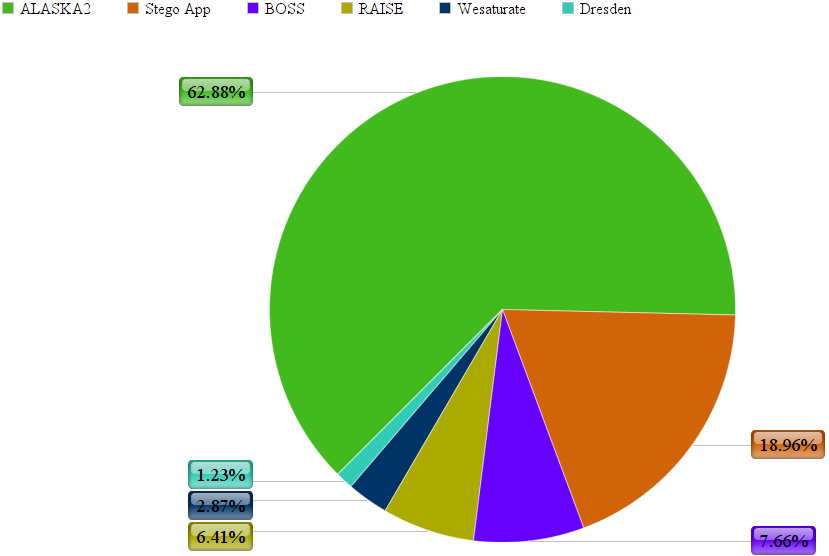}
	
	\caption{Origin of RAW images in the LSSD database.}
	\label{img:general_data:proportion}
\end{figure}

The LSSD database gathers RAW images available on the Internet, mostly from the Alaska\#2 database \cite{Cogranne2019_Alaska} \footnote{Website of the ALASKA challenge\#2: \url{https://alaska.utt.fr/}

Download page : \url{http://alaska.utt.fr/ALASKA_v2_RAWs_scripts.zip}. \label{fn:base:alaska2}} to which we added images from the BOSS \cite{Bas2011-BOSS} \footnote{Challenge BOSS : \url{http://agents.fel.cvut.cz/boss/index.php?mode=VIEW\&tmpl=about}

Download page : \url{ftp://mas22.felk.cvut.cz/RAWs} \label{fn:base:boss}}, RAISE \cite{RAISE_DB} \footnote{Obsolete download link \url{http://mmlab.science.unitn.it/RAISE/}. \label{fn:base:raise}}, Dresden \cite{Gloe:2010aa}\footnote{ \url{http://forensics.inf.tu-dresden.de/ddimgdb}

Download page: \url{http://forensics.inf.tu-dresden.de/ddimgdb/selections}. \label{fn:base:dresden}}  and Wesaturate \footnote{Site closed on February 17, 2020 : \url{http://wesaturate.com/}. \label{fn:base:wesaturate}} databases, as well as StegoApp sites \cite{StegoAppDB}\footnote{\url{https://data.csafe.iastate.edu/StegoDatabase/}. \label{fn:base:stegoapp}}. A total of 127,420 RAW images were collected. Table \ref{tab:intro:database_origins} lists the origin of the RAW images, while Figure \ref{img:general_data:proportion} represents their distributions.

Note that the Alaska\#2 database \cite{Cogranne2019_Alaska} covers  a large variety of ISO parameters, ranging from $20$ (used in general for smartphones) to $51,200$ (only used for high-end devices). Among the $80,005$ images of this database, $11,615$ images have an ISO above $1,000$ while $12,497$ have an ISO below $100$.

The majority of the databases reported in Table \ref{tab:intro:database_origins} are classically used by the community for steganalysis purpose. By combining them, we increase diversity and move to a more "real world" scenario \cite{Ker2013_RealWorld}. We are thus closer to the \enquote{into the wild} spirit \cite{Cogranne2019_Alaska}. The ultimate goal is to reach the diversity findable when browsing the public images of the Internet and social networks.
	\subsection{The \enquote{development} pipeline}
	\label{ssec:dev}
For the Alaska\#2 competition, scripts were used to develop all the images according to some parameters \cite{Cogranne2019_Alaska}. It is thanks to these scripts that it was possible to obtain such a great diversity in the competition by playing on many parameters (see Table \ref{tab:creation:params}). 

We apply these scripts to all the RAW images and we developed them into colour images (ppm format) whose size are $1024\times1024$ pixels (or slightly bigger). If the original image dimensions (width and height) are not equal, the colour image is cropped by taking its central part to get a $1024 \times 1024$ pixels image. A grey level image version is also generated using the standard luminance formula, transforming a RGB colour vector to a scalar representing the grey level:
\begin{equation}
	grey\_value = 0.2989 \times R + 0.5870 \times G + 0.1140 \times B, \nonumber
\end{equation}
where $ R,G,B \in [0, \ldots, 255] $ are the intensities of the red, green and blue channels. We will discuss in the next subsection the development parameters which were used.

As we want to get 2 million images, we add a process to multiply by 16 the number of images. Each colour (respectively the grey-level) ppm image is divided into 16 small images of size $256 \times 256$ pixels. Then, we run a compression of those 16 images, using the standard JPEG quantization matrices, with a quality factor of 75. The compression was carried out using the Python Imaging Library (PIL or \textit{Pillow})\footnote{Documentation: \url{https://pillow.readthedocs.io/en/stable/}.\label{fn:pillow}} package, version 1.1.7, which uses the plugin \enquote{JpegImagePlugin} to compress the images in the format $ 4:4:4 $.

\begin{figure}[!h]
	\centering
	
	\includegraphics[scale=0.45]{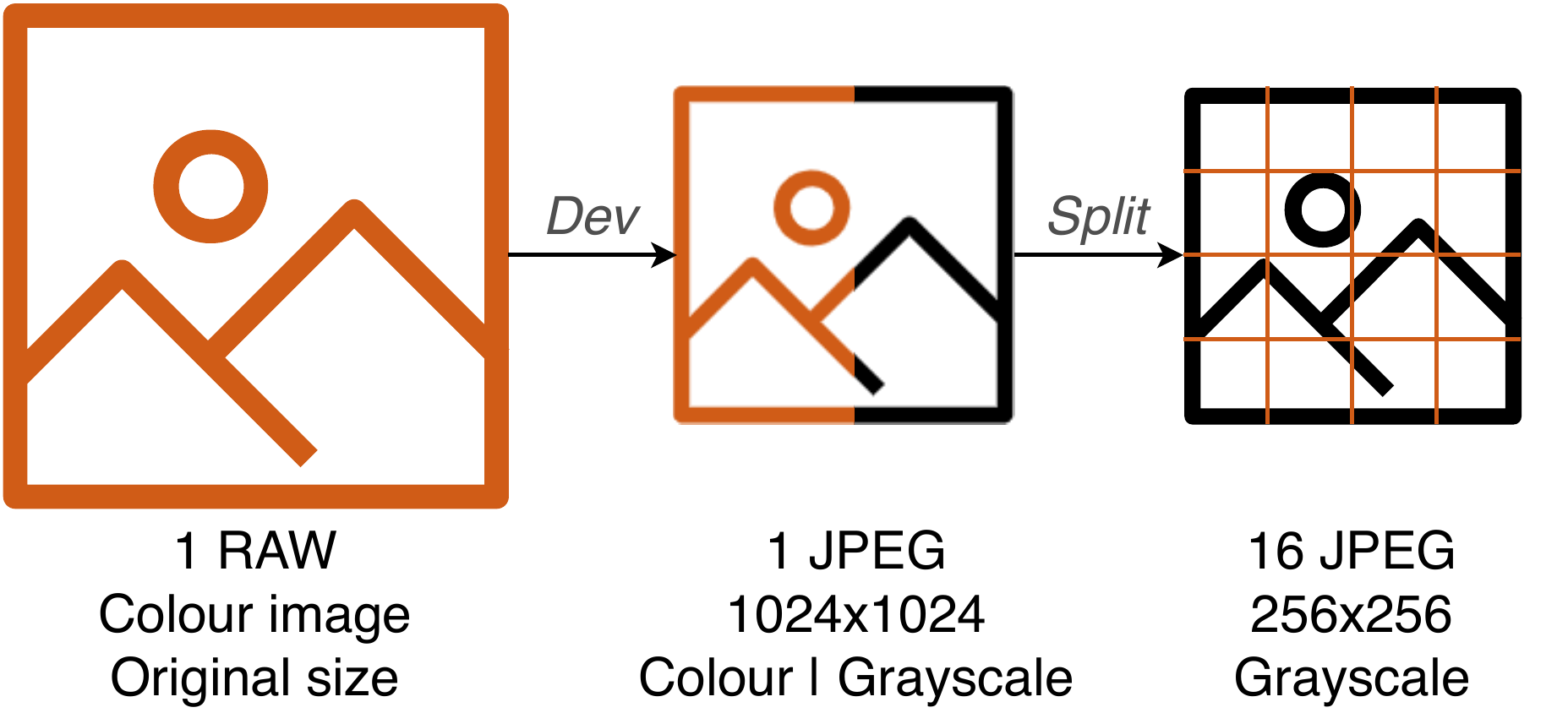}
	
	\caption{Development process of a RAW image.}
	\label{img:dev:process}
\end{figure}

Figure \ref{img:dev:process} schematizes the steps of the complete development process. 

\begin{figure}[h!]
	\centering
	\begin{subfigure}[t]{0.45\textwidth}
		\centering
		\includegraphics[scale=0.6]{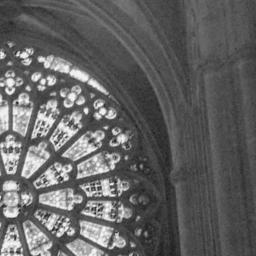}
		\caption{A developed image from the ALASKA database (number 3786).}
	\end{subfigure}%
	~ 
	\begin{subfigure}[t]{0.45\textwidth}
		\centering
		\includegraphics[scale=0.6]{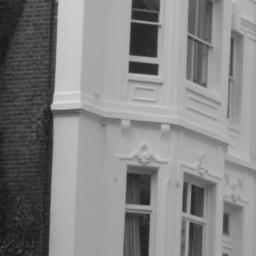}
		\caption{A developed image from the BOSS database (number 6456).}
	\end{subfigure}
	\caption{Two $ 256 \times 256 $ grey-scale images, after development process of the LSSD database.}
	\label{fig:imagesLSSD_grey}
\end{figure}

\begin{figure}[h!]
	\centering
	\begin{subfigure}[t]{0.45\textwidth}
		\centering
		\includegraphics[scale=0.6]{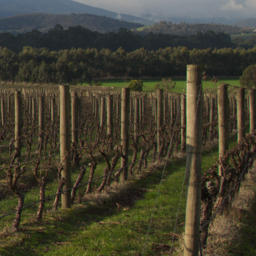}
		\caption{A developed image from the ALASKA database (number 51336).}
	\end{subfigure}%
	~ 
	\begin{subfigure}[t]{0.45\textwidth}
		\centering
		\includegraphics[scale=0.6]{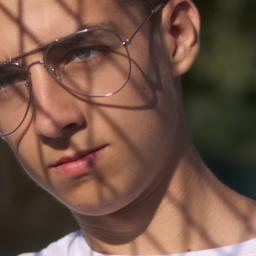}
		\caption{A developed image from the Wesaturate database (index ZYlVRQYDWE).}
	\end{subfigure}
	\caption{Two $ 256 \times 256 $ colour-scale images, after development process of the LSSD database.}
	\label{fig:imagesLSSD_color}
\end{figure}

The $256\times256$ images have semantic content, a resolution and brightness variations that are close from those of images usually processed in steganalysis. Figure \ref{fig:imagesLSSD_grey} (resp. Figure \ref{fig:imagesLSSD_color}) shows two examples of JPEG grey-level (resp. colour) $256\times256$ images of the LSSD.

It takes just under two days for all colour and grayscale images to be developed. 	The users of the LSSD database is free to either download the RAW images and redevelop those images or directly use the colour or grey-level JPEG $256\times256$ images. 

	\subsection{Development parameters}
	\label{sec:pipeline:dev}
In order to obtain the most realistic database, we have tested many development parameters (resize, crop, denoising, quality factor\ldots). Table \ref{tab:creation:params} summarizes all the parameters used during the generation of the dataset, reaching almost two million images.

All the processes explained below are done by using the {\it Rawtherapee}\footnote{Software available at: \url{http://rawtherapee.com}\\ More information can be found at: \url{http://rawpedia.rawtherapee.com}} v5.8 software which is a free, cross-platform raw image processing program. 

\begin{table}[!h]
	\centering
	\caption{Parameters used in the image development process.}
	\label{tab:creation:params}
	\begin{tabular}{|c|c|c|}
		\hline
		\textbf{Number} & \textbf{Name} & \textbf{Value}\\
		\hline
		\hline
		\textit{1} & Demosaicking & Fast or DCB\\
		\hline
		\multirow{6}{*}{\textit{2}} & Resize \& Crop & Yes\\
		& Taille (resize) & $ 1024 \times 1024 $\\
		& Kernel (resize) & \begin{tabular}{@{}c@{}}Nearest (0.2) \\ Bicubic (0.5) \\ Bilinear (0.3)\end{tabular}\\
		& Resize factor & depends on initial size\\
		\hline
		\textit{3} & Unsharp Masking & No\\
		\hline
		\multirow{3}{*}{\textit{4}} & \begin{tabular}{@{}c@{}}Denoise \\ \textit{(Pyramid Denoising)}\end{tabular} & Yes\\
		 & Intensity & [0; 60]\\
		 & Detail & [0; 40]\\
		\hline
		\textit{5} & Micro-contrast & Yes ($ p=0.5 $)\\
		\hline
		\textit{6} & Colour & No\\
		\hline
		\textit{7} & Quality factor & 75\\
		\hline
	\end{tabular}
\end{table}

\textbf{1. Demosaicing:} Demosaicing\ is the process of reconstructing a full-resolu\-tion colour image from the sampled data acquired by a digital camera that applies a colour filter array to a single sensor (see, for example, the overview \cite{menon_2011}).  In the {\it Rawtherapee} software, we can find many demosaicing methods\footnote{Documentation about the different mosaicking methods of {\it Rawtherapee} can be found at:  \url{https://rawpedia.rawtherapee.com/Demosaicing}}. We selected the {\it DCB} method which produces similar results to the best method (AMaZE), plus the {\it Fast} method, based on nearest-neighbor interpolation, which is a lower-quality but very fast method. Notice that another available method called IGV is known to produce the most challenging images to steganalyze \cite{Cogranne2019_Alaska}. For each image, we select either the Fast or the DCB demosaicing algorithm with a probability of 35\% and 65\%, respectively. Demosaicked images are then saved in 16-bit TIFF format using the Python library \textit{Pillow}\textsuperscript{\ref{fn:pillow}}.

\textbf{2. Resize \& Crop:} the image is proportionally resized to final dimensions which are closest to $1024 \times 1024$ pixels as we will divide the resulting image into 16 small images thereafter. If the image is not square, then we crop its center part, assuming that we will keep its semantic content. Resizing is performed using different kernels mentioned in Table 1, and it would have also been possible to use the $8 \times 8$ Lanczos filters.

\label{sec:creation:dev:usm}
\textbf{3. Unsharp Masking (USM):} After resizing, the image can eventually be sharpened. The USM process allows increasing the apparent acutance (edge contrast) of an image, making it appear clearer, even though it technically does not really sharpen the image. This process can be disabled. Note that for the {\it learning} database, the USM has been switched off. USM can be switched on for the development of the {\it test} database; this to introduce strong cover-source mismatch.  More information about USM can be found in Appendix A.1 of \cite{Cogranne2019_Alaska}.


\textbf{4. Denoising:}
When the USM is switched off, denoising is systematically performed using a Pyramid Denoising based on wavelet decomposition. The denoising {\it intensity} parameter follows a gamma distribution with the pdf $ P(x,a) = 10 \times \frac{x^{a-1}\exp(-x)}{\Gamma(a)} $, with $a=4$, and rectified to belongs to $ [0, 100]$. It controls the power of the noise reduction. The {\it Detail} parameter follows a uniform distribution $ \mathcal{U}(\{0..60\}) $, and it controls the restoration of textures in the image due to excessive noise. More information about Pyramid Denoising can be found in \cite{Cogranne2019_Alaska}.

\textbf{5. Micro-contrast:}
Since USM can generate artefacts, it is possible to apply a micro-contrast process. The micro-contrast process is performed after denoising with a probability of $p=0.5$. This process is controlled by two parameters. The {\it strength} parameter follows a gamma distribution with the pdf $ P(x,a) = 100 \times 0.5 \times \frac{x^{a-1}\exp(-x)}{\Gamma(a)} $, with $a=1$, rectified on $ [0, 100] $. This parameters allows to change the strength of the sharpness. The other parameter is the {\it uniformity} for the microcontrast enhancement. The uniformity follows the law $ \lfloor \mathcal{N}(30, 5) \rfloor $ rectified on $ [0, + \infty[ $. That information is recalled in Appendix A.3 \cite{Cogranne2019_Alaska}.
	\subsection{Choice of the JPEG quality factor}
The Quality Factor (QF) of JPEG images is an essential element in the development pipeline. This factor can vary between 0 and 100, with 0 being a very poor quality, 50 being the minimum for good quality and 100 being the best possible. These quality factors are associated with $8\times 8$ quantization matrices that are used in DCT image compression. There are typical (standard) matrices used in JPEG, but it is also possible to design ad-hoc quantization matrices (non-standard). In our case, we only use standard matrices; it results in a lower diversity compared to databases such as ImageNet \cite{imagenet_cvpr09}. Nevertheless, in future work, we would like to integrate this diversity to get as close as possible to real-world images and use image databases like ImageNet.
	\subsection{Reflection about quantization matrix diversity}
For LSSD, we chose the quality factor $Q=75$ (see table \ref{img:dev:process}) with a standard quantization matrix. If we would use a JPEG database such as ImageNet \cite{imagenet_cvpr09}, and desired to generate a database with a $Q$ "around" $75$, we would have to recompress all the images with a factor "equivalent" to $Q=75$. In that case, it is possible to assume that a majority of JPEG images have a factor greater than $75$. By recompressing at a lower factor, we would not introduce recompression artefacts. This new uncontrolled \enquote{real world} base would exhibit statistics of natural JPEG images (i.e. not recompressed), which would resemble those of LSSD, and the performances obtained could be compared with our \enquote{controlled} LSSD base.

Note that it is possible to recover, in the EXIF metadata, the quantization matrices (for each, Y, Cr, Cb, channel) of each JPEG image. With this information, it is easy to identify whether the matrices are standard or non-standard. As recalled in the article of Yousfi and Fridrich \cite{Fridrich2020_Scalable_JPEG_Steg}, the standard formula for obtaining a quantization matrix whatever the channel, given the Quality Factor, $Q$, is:
\begin{equation}
	\textbf{q}(Q) =
	\begin{cases}
		\max \left\{ 
			\textbf{1}, \mbox{round} \left(2 \left( 1 - \frac{Q}{100} \right) \cdot \textbf{q}(50) \right)
		\right\} \ if \ Q>50
		\\
		\min \left\{\{ 
			255 \cdot \textbf{1}, \mbox{round} \left(\frac{50}{Q} \cdot \textbf{q}(50) \right) 
		\right\} \ \ \ \ \ if \ Q \leq 50,
	\end{cases}
	\label{eq:standard_quantizationmatrix}
\end{equation}
$ \textbf{q}(50) $ being the standard quantization matrix for $Q=50$. 

Re-compressing a JPEG image (coming from ImageNet) to a Quality Factor "close" to $Q=75$, can be done by first computing the $\textbf{q}(75)$, and then re-compress the input JPEG image using the $\textbf{q}(75)$.

In the case of a non-standard JPEG input image, if we apply this process, we are losing the quantization diversity. An approach that would preserve this quantization diversity would be to find non-standard matrices noted $\textbf{q}^{(ns)}(75)$, for the re-compression.

To do that, on can first estimate the non standard Quality Factor $Q^{(ns)}$ of the input JPEG image (trough iterative tests using the distance defined in Equation 8 of \cite{Fridrich2020_Scalable_JPEG_Steg}), then compute the $\textbf{q}^{(ns)}(50)$ by multiplying the quantization matrices by the pre-computed "passage"matrices, from $\textbf{q}(Q^{(ns)})$ to $\textbf{q}(50)$, and finally re-use equation \ref{eq:standard_quantizationmatrix} with substituting $\textbf{q}(50)$ by $\textbf{q}^{(ns)}(50)$, this in order to obtain $\textbf{q}^{(ns)}(75)$.

The creation of an ImageNet database re-compressed to Q=75 is postponed to future work.


In conclusion, the diversity of the LSSD can be increased by increasing the development parameters range, by using additional development algorithms, by using various quality factors, and by using non-standard JPEG quantization matrices. Besides, in practice, the diversity of a JPEG colour image can also be increased compared to grey-scale images by using the following various formats: $ 4:4:4 $, $ 4:2:2 $, $ 4:2:0 $ or $ 4:1:1 $.

	\section{Application to DL-based steganalysis}
	\label{sec:steganalysis}
In image classification, a field in which steganalysis is included, it is necessary to learn the neural network used on a training database and then observe its performance on a test database. The images in these bases must absolutely be distinct. The interest of distinguishing these bases is to verify that the network is capable of learning and generalizing the information from the training base to get the best performance from images that it has never analyzed.

The article of Giboulot {\it et al.} \cite{Giboulot2020_Roots}, studying the effects of Unsharp Masking (see in section \ref{sec:creation:dev:usm}), pointed out that USM creates a strong mismatch phenomenon when used in the test set. For this reason, we decided to remove this processing when creating the learning database. The users can thus create a test database, using the USM process, and thus allowing the creation of cover-source mismatch phenomenon, that could be used in order to evaluate the impact of cover-source mismatch on steganalysis. Note that we also suppressed a few other processes such as some demosaicing algorithms and some resizing kernels when creating the learning database.
		\subsection{Training database construction}
		\label{sec:creation:dev:learn_tst}
The RAW database consists of 127,420 images (see Table \ref{tab:intro:database_origins}). We want to generate (from the RAW database) many learning datasets of different sizes from ten thousand to two million grey-scale JPEG images. One possible use is for evaluating the scalability of a steganalysis network, as in \cite{Ruiz2021_Scaling}. It is also necessary to set up a test dataset that will be the same for all learning datasets. This test dataset must be large enough to represent the diversity of developments, without being too disproportionate to the various sizes of the learning datasets. However, it should not be too large, to avoid high computational times in the test phase, even though during the test phase, calculations are faster.

We thus create several training datasets by, recursively, extracting a given number of images from the most extensive database (two million images). In total, we have six different sizes: 10k, 50k, 100k, 500k, 1M, 2M of cover images. So, when a database is used, it is important to take into account the corresponding stego images which doubles the total number of images. For example in the basis \enquote{LSSD\_10k} there are 10,000 cover and 10,000 stego for a total of 20,000 images. In order to clearly identify the impact of increasing the size of the learning set, the smallest bases are included in the largest ones: $10k \subset 50k \subset 100k \subset 500k \subset 1M \subset 2M$. Each database tries to respect at best the initial ratio of the RAW images, which are shown in Table \ref{tab:creation:prop:trn}.

\begin{table}[!h]
	\centering
	
	\caption{Different LSSD database ratio with respect to the initial RAW image ratio.}
	\label{tab:creation:prop:trn}
	
	\begin{tabular}{|c|c|c|c|c|}
		\hline
		\textbf{Base name} & RAW & 100k-2M & 50k & 10k\\
		\hline
		\hline
		ALASKA2 & 62.75\% & = & = & +0.01\% \\
		\hline
		BOSS & 7.84\% & = & = & +0.01\% \\
		\hline
		Dresden & 1.23\% & = & = & +0.01\% \\
		\hline
		RAISE & 6.40\% & = & = & = \\
		\hline
		Stego App DB & 18.92\% & = & = & = \\
		\hline
		Wesaturate & 2.86\% & = & -0.01\% & -0.03\% \\
		\hline
	\end{tabular}
\end{table}
	\subsection{Test database creation}
	\label{sec:creation:test}
We chose to generate a test set of one hundred thousand images. To this end, we isolated 6,250 RAW images with a distribution almost identical to the RAW image database (see section \ref{ssec:raw}). These images will then undergo the same development as the one shown in Table \ref{tab:creation:params}. Note that this RAW test dataset, which is isolated from the training database, allows generating several different test datasets uncorrelated with the JPEG grey-scale image training dataset. Indeed, it is possible to use other development types, with different parameters, to introduce more or less mismatch. In particular, it is possible to incorporate the USM, which produces a strong mismatch and has a significant impact on network performance during the test phase \cite{Giboulot2020_Roots}.

\begin{table}[!h]
	\centering
	\caption{Images distribution of the original  database in the test set.}
	\label{tab:creation:base_tst}
	
	\begin{tabular}{|c|c|c|c|}
		\hline
		\textbf{Database name} & \textbf{Number of images} & \textbf{Percentage} & \textbf{RAW}\\
		\hline
		\hline
		ALASKA2 & 3 970 & 63.52\% & 62.75\%\\
		\hline
		Stego App DB & 1 197 & 19.15\% & 18.92\%\\
		\hline
		BOSS & 496 & 7.94\% & 7.84\%\\
		\hline
		RAISE & 404 & 6.46\% & 6.40\%\\
		\hline
		Wesaturate & 183 & 2.93\% & 2.86\%\\
		\hline
		Dresden & 0 & 0\% & 1.23\%\\
		\hline
	\end{tabular}
\end{table}

Table \ref{tab:creation:base_tst} lists the number of images and the percentage of each database used to form the shared test dataset. Images from Dresden have not been included in order to create a weak \enquote{mismatch} between learning and testing datasets. This phenomenon can be likened to a \enquote{real world} behaviour when the network learns on images that may not be seen again during the test phase.
	\subsection{Format of images}
This database was used to make a test on scalability of a network in \cite{Ruiz2021_Scaling}. In this work, we applied the algorithm J-UNIWARD developed by Holub {\it et al.} \cite{Holub2014_UNIWARD} with a payload of 0.2 bpnzacs (bits per non zero AC coefficients). When Deep Learning is used, it is not possible to give images to the network in JPEG format, so they must be decompressed in MAT format.

Decompressed images are nevertheless much larger than the JPEG images. For example, for a $ 256 \times 256 $ grey-scale image, its size is slightly more than 500 kB because it is stored in double format (the decompressed version is not rounded). Then, a database with almost four million images (cover and stego) takes more than 2 TB. When all data are combined (RAW images, JPEG colour cover, JPEG grey cover, JPEG Grey stego, MAT grey cover and MAT grey stego), we get almost 13 TB of data!

	\section{Conclusion}
The main goal of this work is to provide to the community many controlled databases and a methodology adapted to steganalysis that allows learning on a large scale to get closer to real-world images diversity. The LSSD basis is available on the following website: \url{http://www.lirmm.fr/~chaumont/LSSD.html}

It is already possible to identify the first technical challenges when it comes to processing millions of images, such as the embedding time, the storage space required for a decompressed base. Furthermore, it is required to have scripts significantly optimized to create a new database; otherwise, these times quickly become excessive.

This new public repository gives the community many tools in order to better control their learning. The databases made of few thousand to multiple millions of images, already developed or re-developable is unique in the field. Moreover, the LSSD website is freely accessible, and additionally stores famous RAW databases for conservation since almost half of the RAW images present on the website are no longer downloadable on the Internet. By putting this new database online, it offers the community the possibility to diversify and broaden their research as they wish.

Note that we also generated the colour JPEG images, and those are also downloadable on our website. The studying of colour steganography and colour steganalysis is indeed a hot topic which has recently been addressed during Alaska\#1 \cite{Cogranne2019_Alaska} \cite{Yousfi2019_BreakingALASKA} and Alaska\#2 \cite{Cogranne2020_Alaska2}, \cite{Yousfi2020_Alaska2} \cite{Chubachi2020_Alaska2}.


	\section*{Acknowledgment}
The authors would like to thank the French Defense Procurement Agency (DGA) for its support through the ANR Alaska project (ANR-18-ASTR-0009). We also thank IBM Montpellier and the Institute for Development and Resources in Intensive Scientific Computing (IDRISS/CNRS) for providing us access to High-Performance Computing resources.


%
%
\bibliographystyle{splncs04}
\bibliography{bibliography}
\end{document}